\documentclass[twocolumn,prl,showpacs]{revtex4}

\usepackage{dcolumn}
\usepackage{amsfonts}
\usepackage{amsmath}
\usepackage{amssymb}
\usepackage{bm}

\newif\ifpdf
\ifx\pdfoutput\undefined
\pdffalse 
\else
\pdfoutput=1 
\pdftrue \fi  \ifpdf
\usepackage[pdftex]{graphicx}
\else
\usepackage{graphicx}
\fi

\begin{document}

\ifpdf \DeclareGraphicsExtensions{.jpg,.pdf,.tif} \else
\DeclareGraphicsExtensions{.eps,.jpg} \fi

\newcommand{\brm}[1]{\bm{{\rm #1}}}
\newcommand{\tens}[1]{\underline{\underline{#1}}}
\newcommand{\mm}{\overset{\leftrightarrow}{m}}
\newcommand{\xv}{\bm{{\rm x}}}
\newcommand{\Rv}{\bm{{\rm R}}}
\newcommand{\uv}{\bm{{\rm u}}}
\newcommand{\nv}{\bm{{\rm n}}}
\newcommand{\Nv}{\bm{{\rm N}}}
\newcommand{\ev}{\bm{{\rm e}}}

\title{Phase transitions and soft elasticity of smectic elastomers}

\author{Olaf Stenull}
\affiliation{Department of Physics and Astronomy, University of
Pennsylvania, Philadelphia, PA 19104, USA }

\author{T. C. Lubensky}
\affiliation{Department of Physics and Astronomy, University of
Pennsylvania, Philadelphia, PA 19104, USA }

\vspace{10mm}
\date{\today}

\begin{abstract}
\noindent Smectic-$C$ elastomers can be prepared by
crosslinking, e.g., liquid crystal polymers, in the
smectic-$A$ phase followed by a cooling through the
smectic-$A$ to smectic-$C$ phase transition. This
transition from $D_{\infty h}$ to $C_{2h}$ symmetry
spontaneously breaks rotational symmetry in the smectic
plane as does the transition from a smectic-$A$ to a
biaxial smectic phase with $D_{2h}$ symmetry.  We study
these transitions and the emergent elasticity of the
smectic-$C$ and biaxial phases in three related models and
show that these phases exhibit soft elasticity analogous to
that of nematic elastomers.
\end{abstract}


\maketitle

\begin{figure}
\centerline{\includegraphics[width=8.4cm]{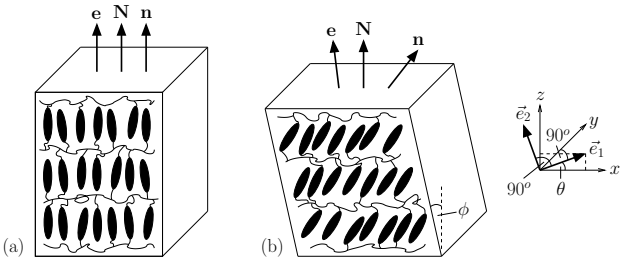}}
\caption{Sample distortion and rotations of the Frank director $\nv$,
the uniaxial anisotropy axis $\ev$, and the layer normal
$\Nv$ in a transition from a (a) Sm$A$ to (b)  a sheared
Sm$C$ elastomer. In this geometry, the the smectic layers
do not rotate.  The figure corresponds to a shear strain
$\gamma \approx u_{xz}^0 <0$ (see text).}
\label{fig1}
\end{figure}

\noindent Liquid crystalline
elastomers~\cite{WarnerTer2003} are remarkable materials
that combine the orientational and positional order of
liquid crystals~\cite{deGennesProst93_Chandrasekhar92} with
the elastic properties of rubber.  The traditional liquid
crystalline nematic, cholesteric, smectic-$A$ (Sm$A$),
smectic-$C$ (Sm$C$), and smectic-$C^*$ phases all exist
\cite{WarnerTer2003} in elastomeric forms. In this letter
we investigate the properties of Sm$C$ and biaxial phases
in elastomers formed via spontaneous symmetry breaking from
Sm$A$ or uniaxial phases and, within mean-field theory, the
phase transitions to them. We introduce and analyze
phenomenological models for these transitions involving
strains only and a model involving strains and the Frank
director specifying the direction of local molecular order.
Our primary result is that monodomain samples of the
emerging biaxial and Sm$C$ phases (see Fig.~\ref{fig1}),
both of which break the continuous rotational symmetry in
the smectic layers, exhibit soft elasticity characterized
by the vanishing of certain elastic moduli and the
associated absence of restoring forces to strains along
specific symmetry directions. As in monodomain nematic
elastomers~\cite{golubovic_lubensky_89,FinKun97,LubenskyXin2003}
this soft elasticity is a consequence of the Goldstone
theorem that requires any phase with a spontaneously broken
continuous symmetry to have modes whose energy vanishes
with wavenumber.

To keep our discussion as simple as possible, we will
consider only smectic elastomers crosslinked in the Sm$A$
phase so that the smectic layers are locked to the
crosslinked matrix \cite{lubensky&Co_94}. Macroscopically,
these Sm$A$ elastomers are simply uniaxial rubbers (or
solids) with $D_{\infty h}$ symmetry, and we will treat
them as such, though we will distinguish between the normal
$\Nv$ to smectic layers and the direction of uniaxial
anisotropy $\ev$.

We employ the usual Langrangian
formalism~\cite{Lagrange-elas} in which mass points in the
undistorted medium, which we take as the reference space,
are labelled by vectors $\xv$. Mass points of the distorted
medium are at positions $\Rv(\xv) = \xv + \uv(\xv)$ in
physical space, which we call the target space. Distortions
of the reference medium are described by the Cauchy
deformation tensor $\tens{\Lambda}$ with components
$\Lambda_{ij} =
\partial R_i /\partial x_j \equiv
\partial_j R_i$, $i,j = x,y,z$. Lagrangian elastic energies are expressed in terms
of the nonlinear strain tensor $\tens{u}$ with components
$u_{ij} ( \xv ) =\frac{1}{2}(\Lambda^T_{ik}\Lambda_{kj} -
\delta_{ij})= \frac{1}{2} (\partial_i u_j +\partial_j u_i +
\partial_i u_k\partial_j u_k)$, which transforms as a
tensor in the reference space and which is invariant with
respect rotations in the target space.  The elastic free
energy density of a uniaxial elastomer to harmonic order in
strains can be expressed as
\begin{align}
\label{uniEn}
f_{\text{uni}} &= \textstyle{\frac{1}{2}} \, C_1\, u_{zz}^2
+ C_2 \, u_{zz} u_{ii} +  \textstyle{\frac{1}{2}} \, C_3\,
u_{ii}^2
\nonumber \\
&+ C_4 \, \hat{u}_{ab}^2 + C_5 \,u_{az}^2,
\end{align}
where we use the Einstein convention on repeated indices
and where $z$ is along the uniaxial axis and indices at the
beginning of the alphabet $a,b,..$ run over $x$ and $y$
only. The tensor
$\hat{u}_{ab}=u_{ab}-\frac{1}{2}\delta_{ab} u_{cc}$ is the
two-dimensional symmetric, traceless strain tensor. The
elastic constant $C_1$ describes dilation or compression
along $z$. $C_4$ and $C_5$ respectively describe shears in
the plane perpendicular to the anisotropy axis and in the
planes containing it. Relative volume change $\delta V/V =
\det \tens \Lambda -1$ can be approximated by $u_{ii}$ in
our Landau expansion in powers of $u_{ij}$, and the
incompressible limit relevant to most elastomers
corresponds to $C_3 \rightarrow \infty$.

The harmonic free energy $f_{\text{uni}}$ describes any
uniaxial solid or elastomer, including Sm$A$ elastomers.
However, to provide a complete description of Sm$A$ and
Sm$C$ elastomers, we need to add terms that describe the
smectic layers and the Frank director and their
interactions with each other and with the elastic medium.
Before adding these terms, we will first investigate phase
transitions in uniaxial elastomers in which either of the
elastic constants $C_4$ or $C_5$ go to zero. The vanishing
of $C_4$ and $C_5$ lead, respectively, to phases with
$D_{2h}$ (orthorhombic) and $C_{2h}$ (triclinic) symmetry.
We will then consider the smectic layers and director and
show that the transition to the Sm$C$ phase, which has
$C_{2h}$ symmetry, is in fact identical to the one in which
$C_5$ goes to zero.

{\em Strain-only theory for $C_4 \to 0$}: If $C_4$ or $C_5$
become negative, as they will in response to an instability
toward biaxial or Sm$C$ ordering of constituent mesogens,
terms higher order in strains must be added to
Eq.~(\ref{uniEn}) to ensure mechanical stability. If $C_4$
becomes negative, order of the shear strain $\hat{u}_{ab}$
sets in and higher order terms featuring $\hat{u}_{ab}$
have to be added which leads to the model elastic energy
\begin{align}
\label{uniEn1}
f_{\text{uni}}^{(1)} = f_{\text{uni}} + A_1  \, u_{zz}
\hat{u}_{ab}^2 + A_2 \,  u_{ii} \hat{u}_{ab}^2 + B \,
(\hat{u}_{ab}^2)^2 ,
\end{align}
where we have dropped qualitatively inconsequential higher
order terms.  It will be useful in the analysis that
follows to regroup the terms in $f_{\text{uni}}^{(1)}$ by
completing the squares in $\frac{1}{2} C_1 u_{zz}^2 + A_1
u_{zz} \hat{u}_{ab}^2$, etc., and to reexpress it as a sum
of two terms
\begin{align}
f_{\text{uni}}^{(1,1)}& =  \textstyle{\frac{1}{2}} \, C_1
v_{zz}^2 + C_2\,  v_{zz} v_{ii}
+\textstyle{\frac{1}{2}} \, C_3 \, v_{ii}^2 + C_5 \, u_{az}^2 \\
f_{\text{uni}}^{(1,2)}& =  C_4\, \hat{u}_{ab}^2 + B_R \,
(\hat{u}_{ab}^2)^2 ,
\end{align}
where $v_{zz}= u_{zz}- \alpha \hat{u}_{ab}^2$, $v_{ii} =
u_{ii} - \beta \hat{u}_{ab}^2$, and where $\beta$, which
vanishes as $C_3 \rightarrow \infty$, $B_R$ and $\alpha$,
which do not, are combinations of the coefficients in
$f_{\text{uni}}^{(1)}$. For a given $\hat{u}_{ab}$,
$f_{\text{uni}}^{(1)}$ is minimized when $u_{zz} = \alpha
\hat{u}_{ab}^2$ and $u_{ii} = \beta \hat{u}_{ab}^2$.  The
equilibrium value of $\hat{u}_{ab}$, $\hat{u}_{ab}^0 =
S(c_a c_b - \frac{1}{2} \delta_{ab})$, where $\brm{c}$ is a
unit vector in the $xy$ plane, is determined by minimizing
$f_{\text{uni}}^{(1,2)}$. The result is $S=0$ for $C_4>0$,
and $S = \pm \sqrt{-C_4 / B_R}$ for $C_4 <0$. The
equilibrium strain $\tens{u}^0=
\frac{1}{2}(\tens{\Lambda}^{0T}\tens{\Lambda}^0 -
\tens{\delta})$ is diagonal with components $u_{xx}^0 =
\frac{1}{2} S + \frac{1}{4}(\beta -\alpha) S^2$, $u_{yy}^0
= - \frac{1}{2} S + \frac{1}{4} (\beta -\alpha)S^2$, and
$u_{zz}^0= \frac{1}{2}\alpha S^2$. The new state is biaxial
with $D_{2h}$ symmetry.

To determine the elastic properties of the new state, we
expand $f_{\text{uni}}^{(1)}$ in powers of $\delta \tens{u}
= \tens{u} - \tens{u}^0$. Since the equilibrium values of
$v_{zz}$, $v_{ii}$ and $u_{az}$ are zero, the expansion of
$f_{\text{uni}}^{(1,1)}$ is trivial. The structure of
$f_{\text{uni}}^{(1,2)}$ is identical to that of an $xy$
model, and it has no restoring force in the ordered phase
for $\delta u_{xy}$: $\delta f_{\text{uni}}^{(1,2)} =  B_R
S^2 ( \delta u_{xx} - \delta{u}_{yy})^2$.  Thus it is clear
that $\delta f_{\text{uni}}^{(1)}$ does not depend on
$\delta u_{xy}$ to harmonic order, i.e., the system is soft
with respect to shears in the $xy$ plane of the original
reference material.

The strain $\delta \tens{u}$ describes distortions relative
to the new biaxial reference state measured in the
coordinates of the original uniaxial state. It is
customary, however, to express the elastic energy in terms
of a strain $\tens{u}'= (\tens{\Lambda}^{0T})^{-1} \delta
\tens{u} \, (\tens{\Lambda}^{0})^{-1}$ measured in the
coordinates $x'_i = x_i + u_i^0 = \Lambda^0_{ij} x_j$ of
the new state. $\tens{\Lambda}^{0}$ is not uniquely
determined by $\tens{u}^0$; rotations in the target space
change $\tens{\Lambda}^0$ but do not change $\tens{u}^0$.
In the present case, it is natural to choose
$\tens{\Lambda}^0$ to be diagonal, i.e., not to rotate the
strain strain after the transition. The elastic energy of
the biaxial state to harmonic order in the new strains is
then
\begin{align}
\label{biaxEn}
f_{D_{2h}}^{\text{soft}}  &= \textstyle{\frac{1}{2}} \,
C_{zzzz}\, (u^\prime_{zz})^2 + C_{xzxz} \,
(u^\prime_{xz})^2 + C_{yzyz} \, (u^\prime_{yz})^2
\nonumber \\
&+ C_{zzxx} \, u^\prime_{zz} u^\prime_{xx} + C_{zzyy} \,
u^\prime_{zz} u^\prime_{yy} + \textstyle{\frac{1}{2}} \,
C_{xxxx}\, (u^\prime_{xx})^2
\nonumber \\
&+\textstyle{\frac{1}{2}} \, C_{yyyy}\, (u^\prime_{yy})^2 +
C_{xxyy} \, u^\prime_{xx} u^\prime_{yy},
\end{align}
where the elastic constants depend on the original elastic
constants featured in Eq.~(\ref{uniEn1}) and the order
parameter $S$, and for $S
\to 0$ this energy reduces to the uniaxial
energy~(\ref{uniEn}). If we take the incompressible limit,
$C_3 \to \infty$,  the specifics of the new elastic
constants are affected but the form of the elastic
energy~(\ref{biaxEn}) remains the same.

Because there was no $\delta u_{xy}$ term in the expansion
of $f_{\text{uni}}^{(1)}$, there is no term proportional to
$u_{xy}^{\prime 2}$ as there would be in conventional
orthorhombic systems. Thus, there is no restoring force to
$xy$-stresses, i.e., to opposing forces along $\pm x$
applied to opposite surfaces perpendicular to $y$ or
opposing forces along $\pm y$ applied to opposite surfaces
perpendicular to $x$. In addition, as is the case for
stresses perpendicular to the anisotropy axis in a nematic
elastomer \cite{FinKun97,LubenskyXin2003}, it requires no
stress to stretch the sample along $y$ direction, up to a
critical strain value. The same soft mode and many more
were predicted by Warner and
Kutter~\cite{warner_kutter_2002} for biaxial nematics
forming spontaneously from an isotropic elastomer.

{\em Strain-only theory for $C_5 \to 0$}: When $C_5$ is
driven negative the uniaxial state becomes unstable to
shear in the planes containing the anisotropy axis, and the
uniaxial energy (\ref{uniEn}) must be augmented with higher
order terms involving $u_{az}$ to stabilize the system,
\begin{align}
\label{uniEn2}
f_{\text{uni}}^{(2)} &= f_{\text{uni}} + D_1\, u_{zz}
u_{az}^2 + D_2 \,  u_{ii} u_{az}^2 + D_3 \, \hat{u}_{ab}
u_{az} u_{bz}
\nonumber \\
&+ E  \, (u_{az}^2)^2 ,
\end{align}
where we omitted all unimportant symmetry-compatible higher
order terms. To study the ordered phase of this free energy
when $C_5<0$, we proceed in much the same way as we did for
the biaxial state of $f_{\text{uni}}^{(1)}$. We complete
squares to write $f_{\text{uni}}^{(2)}$ as the sum of two
terms:
\begin{align}
f_{\text{uni}}^{(2,1)}& =  \textstyle{\frac{1}{2}} \, C_1\,
w_{zz}^2 + C_2 \, w_{ii}w_{zz}
+\textstyle{\frac{1}{2}}\, C_3 \, w_{ii}^2 + C_4 \, w_{ab}^2 \\
f_{\text{uni}}^{(2,2)}& =  C_5 \, u_{az}^2 + E_R \,
(u_{az}^2)^2 ,
\end{align}
where $w_{zz} = u_{zz} - \sigma u_{az}^2$, $w_{ii} = u_{ii}
- \tau u_{az}^2$, and $w_{ab} = \hat{u}_{ab} - \omega
(u_{az}u_{bz} - \frac{1}{2}\delta_{ab} u_{cz}^2)$ where
$\sigma$, $\tau$ ($\rightarrow 0$ as $C_3 \rightarrow
\infty$), $\omega$, and $E_R$ are functions of the
parameters in $f_{\text{uni}}^{(2)}$. The equilibrium value
of $u_{az}$ is determined by minimizing
$f_{\text{uni}}^{(2,2)}$, which has $xy$ symmetry.  For
$C_5>0$, $u_{az} = 0$; for $C_5<0$ and order along $x$,
$u_{xz}^0\equiv S = \pm \sqrt{- C_5/(2 E_R)}$, and
$u_{yz}^0=0$. The other components of $\tens{u}^0$ are
$u_{zz}^0 =  \sigma S^2$, $u_{xx}^0 = \frac{1}{2}(\tau +
\omega- \sigma)S^2$, and $u_{yy}^0 = \frac{1}{2}(\tau-
\omega -\sigma) S^2$. Unlike the biaxial case, $\tens{u}^0$
is not diagonal; it has nonvanishing $xz$ and $zx$
components that lead to $C_{2h}$ rather than $D_{2h}$
symmetry. Expanding $f_{\text{uni}}^{(2)}$ in powers of
$\delta \tens{u} = \tens{u} - \tens{u}^0$, we find that
$\delta f_{\text{uni}}^{(2,2)} = 4 E_R S^2 (\delta
u_{xz})^2$ is independent of $\delta u_{yz}$, and we might
naively expect the system to exhibit softness with respect
to $u_{yz}$.  This, however, is not the case because
$f_{\text{uni}}^{(2,1)}$ depends on $\delta u_{yz}$ via the
$C_4$ term: $2C_4 (\delta u_{xy} - \omega S \, \delta
u_{zy})^2$. Thus, the softness of the ordered phase with
$C_{2h}$ symmetry is more subtle than that of the biaxial
phase with $D_{2h}$ symmetry.

To determine the energy of strains relative to the new
ground state, we need to choose how we define our new
coordinate system relative to it.  It is easiest to
visualize the new state as emerging from a simple shear as
shown in Fig.\ \ref{fig1} in which $\Lambda_{xz}^0 =
\partial R_x^0/\partial z \equiv \gamma$ is nonzero but
$\Lambda_{zx}^0 = \partial R_z^0/\partial x = 0$.  In this
case, the only nonzero components of $\tens{\Lambda}^0$ are
the diagonal components and $\Lambda_{xz}^0$ which can be
expressed in terms of $\tens{u}^0$: $\Lambda_{xx}^0=
\sqrt{1 + 2 u_{xx}^0}$, $\Lambda_{yy}^0=\sqrt{1+2
u_{yy}^0}$, $\Lambda_{xz}^0 = \gamma= 2
u_{xz}^0/\Lambda_{xx}^0$, and $\Lambda_{zz}^0=\sqrt{1 + 2
u_{zz}^0 - \gamma^2} $. The elastic energy can now be
written in terms of $\tens{u}'= (\tens{\Lambda}^{0T})^{-1}
\delta \tens{u} \, (\tens{\Lambda}^{0})^{-1}$ as
\begin{align}
\label{C2hEn}
& f_{C_{2h}}^{\text{soft}} = \textstyle{\frac{1}{2}} \, \bar{C} \left[ \cos \theta \,
u^\prime_{xy} + \sin \theta \, u^\prime_{yz} \right]^2 +
\textstyle{\frac{1}{2}} \, C_{zzzz} \, (u^\prime_{zz})^2
\nonumber \\
& + C_{xzxz} \, (u^\prime_{xz})^2 + C_{zzxx} \,
u^\prime_{zz} u^\prime_{xx} + C_{zzyy} \, u^\prime_{zz}
u^\prime_{yy}
\nonumber \\
& + \textstyle{\frac{1}{2}} \, C_{xxxx} \,
(u^\prime_{xx})^2+ \textstyle{\frac{1}{2}} \, C_{yyyy} \,
(u^\prime_{yy})^2 + C_{xxyy} \, u^\prime_{xx} u^\prime_{yy}
\nonumber \\
& + C_{xxxz} \, u^\prime_{xx} u^\prime_{xz}+ C_{yyxz} \,
u^\prime_{yy} u^\prime_{xz} + C_{zzxz} \, u^\prime_{zz}
u^\prime_{xz},
\end{align}
where the angle $\theta$ and the elastic constants
$\bar{C}$, $C_{zzzz}$ and so on depend on the original
elastic constants in Eq.~(\ref{uniEn2}) and $S$ so that one
retrieves the uniaxial energy (\ref{uniEn}) for $S\to 0$.
To lowest order in $S$, $\tan \theta = \omega S$. The
elastic energy~(\ref{C2hEn}) has only 12 (including
$\theta$) rather than the 13 elastic constants of
conventional triclinic solids \cite{triclinic}. There are
only two rather than three independent elastic constants in
the subspace spanned by $u^\prime_{xy}$ and
$u^\prime_{yz}$. With the introduction of $\vec{v}=(u_{xy}^\prime,
u_{zy}^\prime)$, elastic energies in this subspace can be
expressed as $\frac{1}{2} \vec{v} \cdot \mm \cdot \vec{v}$
where $\mm$ is a $2 \times 2$ matrix.  In general $\mm$ has
two independent eigenvalues $m_1$ and $m_2$ with respective
associated orthnormal eigenvectors $\vec{e}_1$ and
$\vec{e}_2$ with respect to which $\mm$ is diagonal:
$\vec{v} \cdot \mm \cdot \vec{v} = m_1 (\vec{e}_1 \cdot
\vec{v})^2 + m_2 (\vec{e}_2 \cdot \vec{v})^2$. The first
terms in Eq.\ (\ref{C2hEn}) is of the form $\frac{1}{2} m_1
(\vec{e}_1 \cdot \vec{v})^2$ with $m_1 = \overline{C}$ and
$\vec{e}_1 = (\cos \theta, \sin \theta)$, and we conclude
that $m_2 = 0$.  Thus distortions along $\vec{e}_2 =
(-\sin\theta, \cos \theta)$, i.e., distortions for which
$\vec{v} \parallel \vec{e}_2$ cost no energy (see
Fig.~\ref{fig1}). Stated differently, there are no
restoring forces to stress $-\sin \theta \, \sigma_{xy} +
\cos \theta \, \sigma_{zy} = e_{2i} \, \sigma_{iy}$, i.e.,
to stress in the $xz$ plane directed along $\vec{e}_2$ or
stress in the plane perpendicular to $\vec{e}_2$ directed
along $y$.

{\em Theory with director and smectic layers}: The
following theory generalizes the achiral limit of a
continuum theory for Sm$C^*$ elastomers by Terentjev and
Warner~\cite{TerentjevWar1994} in a formalism that ensures
invariance with respect to arbitrary rather than
infinitesimal rotations of both the director and mass
points.

In traditional uncrosslinked liquid crystals, there is no
reference space, and all physical fields like the smectic
layer-displacement field $U$, the layer normal $\Nv$, and
the Frank director $\nv$ are defined at real or target
space points $\Rv$ and they transform as scalars, vectors,
and tensors under rotations in the target space. In the
Lagrangian theory of elasticity, fields are defined at
reference space points $\xv$, and they transform into
themselves under the symmetry operations of that space. To
develop a theory of liquid-crystalline elastomers, it is
necessary to combine target-space liquid crystalline fields
and reference space elastic variables to produce scalars
that are invariant under arbitrary rotations in the target
space and under symmetry operations of the reference space.
This requires that we be able to represent vectors and
tensors in either space \cite{LubenskyXin2003}. The matrix
polar decomposition theorem \cite{HornJoh1991} applied to
the Cauchy tensor $\tens{\Lambda}$ provides a route to this
representation. Like any non-singular matrix,
$\tens{\Lambda}$ can be decomposed into the product of an
orthogonal matrix $\tens{O}$ times a symmetric matrix:
$\tens{\Lambda} = \tens{O}\,\tens{M}^{1/2}$, where
$\tens{M}^{1/2}$ is the symmetric square root of the symmetric
matrix $\tens{M} = (\tens{\Lambda}^T \tens{\Lambda})=
(\tens{1}+2 \tens{u})$, which depends only on the symmetric
strain $\tens{u}$, and $\tens{O} = \tens{\Lambda}\,
\tens{M}^{-1/2}$.  The orthogonal matrix $\tens{O}$
converts (or rotates) any reference space vector
$\tilde{\brm{a}}$ to a target space vector $\brm{a}$ via
$\brm{a}=\tens{O} \cdot \tilde{\brm{a}}$ and a target-space
vector to a reference space vector via $\tilde{\brm{a}}=
\tens{O}^T\cdot \brm{a}$. Equipped with $\tens{O}$, we can
rotate the target space director $\nv$ to a reference space
vector $\tilde{\nv}\equiv (\tilde{\brm{c}}, \tilde{n}_z)$,
where $\tilde{n}_z = \sqrt{1 - \tilde{c}_a^2}$, and rotate
the reference space anisotropy vector $\tilde{\brm{e}}=
(0,0,1)$ to a target-space vector $\brm{e} = \tens{O} \cdot
\tilde{\brm{e}}$.  From these, we can form invariant
couplings like $(\tilde{\nv}\cdot \tilde{\brm{e}})^2 = (\nv
\cdot \brm{e})^2 = 1-\tilde{c}_a^2$, $\tilde{c}_a u_{ab}
\tilde{c}_b$, and $\tilde{c}_a u_{az}\tilde{n}_z$ as
building blocks the elastic energy.

To treat smectic layers, we need to discuss in more detail
the smectic displacement field $U$ and the layer normal
$\Nv$.  The smectic mass-density-wave amplitude for a
system with layer spacing $d$ has a phase $\phi ( \Rv) =
q_0 [R_z - U(\Rv)]$ where $q_0 = 2 \pi /d$.  Since there is
a one-to-one mapping from the reference space points $\xv$
to the target-space points $\Rv(\xv)$, we can express
$\phi$ as a function of $\xv$ as $\phi(\xv) = q_0 [ z + u_z
( \xv) - U(\Rv(\xv))]$.  We are only considering systems
crosslinked in the smectic phase.  In these systems, the
smectic mass-density wave cannot translate freely relative
to the reference material, and there is a term in the
free-energy density $\frac{1}{2}A (u_z - U)^2$ that locks
the smectic field $U$ to the displacement field
$u_z$~\cite{lubensky&Co_94}.  In what follows, we will take
this lock-in as given and set $U=u_z$. The layer normal in
the target space is $N_i = \nabla_i \phi/|\bm{\nabla}
\phi|$, where $\nabla_i \phi \equiv
\partial \phi/\partial R_i = \partial_j \phi\Lambda_{ji}^{-1}$. Thus, when
$U$ is locked to $u_z$, $N_i =
[(\tens{M}^{-1})_{zz}]^{-1/2}\Lambda_{zi}^{-1}$,
$\tilde{N}_i =[(\tens{M}^{-1})_{zz}]^{-1/2}M_{zi}^{-1/2}$,
and to harmonic order in $\tilde{c}_a$ and $u_{ij}$,
$\tilde{\Nv}\cdot \tilde{\nv}= \Nv \cdot \nv = 1-
\tilde{c}_a^2 - u_{za} \tilde{c}_a + \cdots$.

We can now develop a full phenomenological free energy for
the Sm$A$--to--Sm$C$ transition in an elastomer. In the
equilibrium Sm$A$ phase, the director is parallel to both
the layer normal $\Nv$ and the anisotropy axis $\ev$, which
are parallel to each other, and there are energy costs
proportional to $(\tilde{\Nv}\cdot \tilde{\nv})^2$ and
$(\tilde{\ev}\cdot \tilde{\nv})^2$ associated with
deviations from this equilibrium.  These combine to yield a
term in the free-energy density proportional to
$\tilde{c}_a^2$ and higher order terms involving the strain
and strain-director coupling. With the addition of higher
order terms in $(\tilde{\Nv}\cdot \tilde{\nv})^2$ and
$(\tilde{\ev}\cdot \tilde{\nv})^2$ to stabilize the Sm$C$
phase, the free energy up to inconsequential higher order
terms becomes
\begin{align}
\label{completeEn}
f = f_{\text{uni}} + f_{\text{tilt}} + f_{\text{coupl}},
\end{align}
where $f_{\text{uni}}$ is the uniaxial energy (\ref{uniEn})
and
\begin{align}
\label{Htilt}
f_{\text{tilt}} &= \textstyle{\frac{1}{2}} \, r \,
\tilde{c}_a^2 +\textstyle{\frac{1}{4}} \, g \,
(\tilde{c}_a^2)^2 ,
\\
\label{Hcoupl}
f_{\text{coupl}} &=  \lambda_1\,  \tilde{c}_a^2 u_{zz} +
\lambda_2 \, \tilde{c}_a^2 u_{bb}  +  \lambda_3 \,
\tilde{c}_a \hat{u}_{ab} \tilde{c}_b
\nonumber \\
&+  \lambda_4 \, \tilde{c}_a u_{az} \tilde{n}_z .
\end{align}
When $\lambda_1=\lambda_2= \lambda_3=0$, this model is
equivalent to that studied in Ref.~\cite{TerentjevWar1994}
when polarization is ignored.

We can now analyze the transition to the Sm$C$ phase in
exactly the same way as we did in the strain only model. We
complete the squares involving the strains and the
director-strain couplings. The result is an expression for
the free energy that is the sum of a term $f^{(1)}$ quadratic
in shifted strains of the form $u_{zz} + \rho
\tilde{c}_a^2$, etc. and a term $f^{(2)}=\frac{1}{2} \, r_R
\, \tilde{c}_a^2 + \frac{1}{2} \, g_R \, (\tilde{c}_a^2)^2$
depending only on $\tilde{c}_a$, where $r_R = r -
\lambda_4^2/(2C_5)$. With $\tilde{\brm{c}}$ chosen to align
along $x$, minimization yields $\tilde{c}_x^0 \equiv S =
\pm \sqrt{-r_R/g_R}$, $u_{xz}^0 = -\lambda_4 S/(2C_5)$ and
diagonal components of $\tens{u}^0$ proportional to $S^2$.
As anticipated from the strain-only model for $C_5 \to 0$,
the Sm$C$ phase is a sheared biaxial phase with $C_{2h}$
symmetry. If $\tilde{c_a}$ is integrated out of $f$, the
result is identical to $ f_{\text{uni}}$ with $C_5$
renormalized to $C_{5,R} = C_5 - \lambda_4^2/(2r)$, which
vanishes at $r_R = 0$.

To address the elastic properties of the Sm$C$ phase within
our last model, we expand the complete elastic energy
(\ref{completeEn}) in terms of  $\delta \tens{u} = \tens{u}
- \tens{u}^0$ and $\delta \tilde{\brm{c}} = \tilde{\brm{c}}
- \tilde{\brm{c}}^0$. Since $f^{(2)}$ has $xy$ symmetry,
its expansion has a $(\delta \tilde{c}_x)^2$ but no
$(\delta \tilde{c}_y)^2$ term. The expansion of $f^{(1)}$
has terms quadratic in $\delta \tens{u}$ and couplings
between $\delta \tens{u}$ and $\delta \tilde{c}_a$.  Since
$\delta \tilde{c}_x$ is massive, it can be integrated out.
Converting the resulting expression to the strain variable
$\tens{u}^\prime$ of the Sm$C$ state, we obtain an elastic
energy that is identical in form to Eq.~(\ref{C2hEn}) but
with an additional term proportional to $[ \delta
\tilde{c}_y + \mu ( u_{xy}^\prime + \zeta u_{yz}^\prime )
]^2$, where $\mu$ and $\zeta$ are dimensionless
coefficients that depend on the original elastic constants
and $S$. This expression shows clearly that $\delta
\tilde{c}_y$ can relax locally to $ - \mu \left(
u_{xy}^\prime + \zeta u_{yz}^\prime \right)$ to eliminate
the dependence of the elastic energy on $u_{xy}^\prime +
\zeta u_{yx}^\prime$ and produce an energy with softness
identical to that of Eq.~(\ref{C2hEn}).

The above analysis provides us with the strain $\tens{u}^0$
and the $c$-director $\tilde{\brm{c}}^0$ in the coordinates of
the reference space without rotation. To visualize what
happens to $\ev$, $\Nv$, and $\nv$ in the Sm$C$ phase, it
is useful to consider the simple shear of Fig.~\ref{fig1}
in which $\tan \, \phi = \Lambda_{xz}^0 = \gamma$, and
$\Lambda_{zx}^0=0$. To lowest order in $S$, $\gamma = 2
u_{xz}^0=-\lambda_4 S/C_5$, and $O_{ij} = \delta_{ij} +
\frac{1}{2} \epsilon_{izj} \gamma$. Then $\nv \approx (n_x,
0,1)$ where $n_x = S+\frac{1}{2}\gamma$, and $e_i = O_{iz}
= (\frac{1}{2} \gamma, 0,1)$.  Thus, at the transition
$\ev$ and $\nv$ rotate relative to $\Nv$ by different
amounts. The layer normal vector $\Nv$, however, remains
fixed in the chosen geometry: $N_i \sim \Lambda_{zi}^{-1}$
only has a nonvanishing $z$ component and is equal to
$(0,0,1)$. Thus, the smectic layers remain parallel to the
$x$-axis as one would intuitively expect in the chosen
geometry.

{\em Concluding remarks}: We have presented models for
transitions from uniaxial Sm$A$ elastomers to biaxial and
Sm$C$ elastomers, and we have calculated the nature of the
soft elasticity, required by symmetry, of monodomain
samples of these phases.  We hope that our work encourages
experiments to probe this soft elasticity.

Support by the National Science Foundation under grant DMR
00-95631(TCL) is gratefully acknowledged.


\begin{thebibliography}{}
\bibitem{WarnerTer2003} For a review see M. Warner and E.M.~
Terentjev, {\em Liquid Crystal Elastomers} (Clarendon
Press, Oxford, 2003).
\bibitem{deGennesProst93_Chandrasekhar92}
For a review see P. G. de~Gennes and J.~Prost, {\em The
Physics of Liquid Crystals} (Clarendon Press, Oxford,
1993); S.~Chandrasekhar,  {\em Liquid Crystals} (Cambridge
University Press, Cambridge, 1992).
\bibitem{golubovic_lubensky_89}
L. Golubovi\'{c} and T. C. Lubensky, Phys. Rev. Lett. {\bf
63}, 1082 (1989); Peter D. Olmsted, J. Phys. II (France) 4,
2215 (1994).
\bibitem{FinKun97}
H. Finkelmann, I. Kundler, E. M. Terentjev, and M. Warner,
J.\ Phys.\ II (France) {\bf 7}, 1059 (1997); G. C. Verwey,
M. Warner, and E. M. Terentjev, J.\ Phys.\ II (France) {\bf
6}, 1273 (1996); M. Warner, \newblock J.\ Mech.\ Phys.\
Solids {\bf 47}, 1355 (1999).
\bibitem{LubenskyXin2003}
T. C. Lubensky, R. Mukhopadhyay, L. Radzihovsky, and X.J.
Xing, Phys. Rev. E {\bf 66}, 011702 (2002).
\bibitem{lubensky&Co_94}
T. C. Lubensky, E. M. Terentjev, and M. Warner, J.\  Phys.\
II (France) {\bf 4}, 1457 (1994). Without this lockin, the
phase of the smectic mass-density-wave can translate freely
relative to the elastomer as it can in smectics in aerogels,
L. Radzihovsky and J. Toner, Phys. Rev. B {\bf 60}, 206
(1999).
\bibitem{Lagrange-elas} L.D. Landau and E.M. Lifshitz, {\it Theory of Elasticity},
3rd Edition (Pergamon Press, New York, 1986); P.M. Chaikin
and T.C. Lubensky, {\it Principles of Condensed Matter
Physics} (Cambridge Press, Cambridge, 1995).
\bibitem{warner_kutter_2002}
M. Warner and S. Kutter, Phys. Rev. E {\bf 65}, 051707
(2002).
\bibitem{triclinic}
See, e.g., N. W. Ashcroft and N. D. Mermin, {\em Solid
State Physics}, (Saunders, Philadelphia, 1976).
\bibitem{TerentjevWar1994}
 E. M. Terentjev and M. Warner,
Journal de Physique II 4, 849 (1994).
\bibitem{HornJoh1991}
See, e.g., R. A. Horn and C. R. Johnson, {\em Topics in
Matrix Analysis} (Cambridge University Press, New York,
1991).
\end{thebibliography}
\end{document}